# The Challenges of Hardware Synthesis from C-like Languages

Stephen A. Edwards*
Department of Computer Science, Columbia University, New York

MANY TECHNIQUES for synthesizing digital hardware from C-like languages have been proposed, but none have emerged as successful as Verilog or VHDL for register-transfer-level design. This paper looks at two of the fundamental challenges: concurrency and timing control.

Familiarity is the main reason C-like languages have been proposed for hardware synthesis. Synthesize hardware from C, proponents claim, and we will be able to turn a C programmer into a hardware designer. Another common motivation is hardware/software codesign: today's systems usually contain a mix of hardware and software, and it is often unclear initially which portions to implement in hardware. Here, using a single language should simplify the migration task.

**C was developed** by Dennis Ritchie in the early 1970s [18], it was derived from BCPL [17]. Both languages' abstractions are close to processor data types and operations. BCPL treats memory as an array of words; integers, pointers, and characters are represented as a word. For the PDP-11, Ritchie added character, integer, and floating-point types. C's arrays are a side effect of its pointer semantics, which enables simple, efficient implementations, but also demands compilers with aggressive optimization to perform costly pointer analysis.

That C has types that match what the processor directly manipulates and pointers instead of a first-class array type is troubling when synthesizing hardware from C. Bit vectors are natural in hardware, yet C only supports four sizes. C's memory model is an undifferentiated array of bytes, yet many small, varied memories are most effective in hardware.

**C-like hardware synthesis languages** have been proposed since the late 1980s (Table 1). Surveys include a longer version of this paper [6] and De Micheli [3].

Stroud et al.'s early Cones [23] synthesized each functions in a combinational block. Its strict C subset handled conditionals; loops, which it unrolled; and arrays treated as bit vectors.

As input for their Olympus synthesis system [4], Ku and De Micheli developed HardwareC [12], a behavioral hardware language with support for hardware structure and hierarchy.

Galloway's Transmogrifier C [8] supports loops, conditionals, and integer arithmetic. It places cycle boundaries at function calls and at the beginning of *while* loops.

The SystemC [9] C++ library supports hardware and system modeling. While most popular for modeling (it provides concurrency with lightweight threads [13]), a subset of the language can be synthesized. Classes model hierarchical structures containing combinational and sequential processes.

In IMEC's Ocapi system [19], the user's C++ program runs to generate a data structure that represents hardware. Supplied classes provide mechanisms for specifying datapaths, finite-state machines, etc. (Paško et al. [16] adds RAM interfaces). The result is translated into a language such as Verilog and synthesized. Lipton et al.'s PDL++ [14] is similar.

*sedwards@cs.columbia.edu      http://www.cs.columbia.edu/~sedwards
Edwards is supported by an NSF CAREER award, a grant from Intel corporation, an award from the SRC, and from New York State's NYSTAR program.

| | |
|---|---|
| Cones [23] | Early, combinational only |
| HardwareC [12] | Behavioral synthesis-centric |
| Transmogrifier C [8] | Limited scope |
| SystemC [9] | Verilog in C++ |
| Ocapi [19] | Algorithmic structural descriptions |
| C2Verilog [21] | Comprehensive; company defunct |
| Cyber [24] | Restricted C with extensions (NEC) |
| Handel-C [2] | C with CSP (Celoxica) |
| SpecC [7] | Resolutely refinement-based |
| Bach C [10] | Untimed semantics (Sharp) |
| CASH [1] | Synthesizes asynchronous circuits |

Table 1: C-like languages/compilers (chronological order).

C2Verilog, developed at CompiLogic (renamed C Level Design then bought by Synopsys in November 2001) has truly broad support for ANSI C. It can translate pointers, recursion, dynamic memory allocation, and other thorny C constructs. Soderman and Panchul [21, 22] hold a broad patent on C-to-Verilog translation [15] describing the compiler.

NEC's Cyber system [24] accepts a C variant dubbed BDL that contains hardware extensions but prohibits recursive functions and pointers. Timing can be implicit or explicit.

Celoxica's Handel-C [2] adds constructs for parallel statements and OCCAM-like rendezvous communication. Each assignment statement runs in one cycle.

Gajski et al.'s SpecC [7] adds constructs for finite-state machines, concurrency, pipelining, and structure through thirty-three keywords [5]. Systems written in the complete language must be refined into the synthesizable subset.

Sharp's Bach C [10] adds explicit concurrency and rendezvous communication. The compiler does the scheduling; the number of cycles taken by each construct is not set by a rule. It supports arrays but not pointers.

Budiu et al.'s CASH [1] is unique because it generates asynchronous hardware. It identifies instruction-level parallelism in ANSI C and generates asynchronous dataflow circuits.

**Concurrency** is the biggest difference between hardware, for which is is fundamental, and software. Efficient software algorithms are rarely the best choice in hardware. More disturbing is that C and C++ are optimized for expressing sequential algorithms and contain no language-level support for concurrency, in part because there is no agreed-upon model for parallel programming [20]. The absence of concurrency support means it must be added or inferred by the compiler.

About half the languages require the programmer to express concurrency with parallel constructs. HardwareC, SystemC, and Ocapi all use process-level constructs; Handel-C, and SpecC can also group concurrent statements. SystemC's parallelism resembles Verilog or VHDL's: a system is a collection of clock-edge-triggered processes. Handel-C, SpecC, and Bach C's approaches are more software-like: their constructs dispatch groups of instructions in parallel.

Concurrency introduces a fundamental change to the language, demanding substantially different programmer thinking. Even if s/he is experienced with concurrent programming with the usual thread-and-shared-memory model, the parallel constructs in hardware languages differ substantially.

Other languages present a sequential model to the programmer and rely on the compiler to identify parallelism. While compilers for languages with parallel constructs also identify parallelism, Cones, Transmogrifier C, C2Verilog, and CASH rely on the compiler completely. Cones flattens each function, including loops and conditionals, into a single two-level networks. CASH, by contrast, takes a VLIW-compiler-like approach, analyzing inter-instruction dependencies and scheduling instructions to maximize parallelism.

Two common approaches to identifying parallelism differ in their granularity. Instruction-level parallelism (ILP) groups nearby instructions that can run simultaneously. Now the preferred approach in the computer architecture community, it seems that ILP beyond about five simultaneous instructions is unlikely due to fundamental limits [25, 26]. Pipelining, the second approach, requires less hardware than ILP but can be less effective. Again, dependencies and control-flow transfers limit parallelism. Pipelining works well on regular loops, e.g., in scientific computation [11], but is less effective in general.

For hardware, relying on the compiler to expose parallelism is awkward because using it effectively requires understanding details of the compiler's operation. Efficient implementations demand careful coding, and appropriate idioms would be awkward for programmers accustomed to writing efficient C.

**Time** is absent from the C programming model. It guarantees causality, but says nothing about execution time. A simple model for both programmers and compilers, it can make achieving timing constraints difficult. The transparency of C software compilation makes gross improvements easy, but improving an already-optimized fragment is difficult.

Meeting a performance target under power and cost constraints is usually mandatory in hardware, since it is always easier to implement a function in software. Thus, any hardware synthesis technique needs a way to meet timing constraints.

The C-like languages in this paper generate synchronous hardware (except Cones, which generates combinational logic, and CASH, which generates asynchronous), so there must be a mechanism for dividing time into clock cycles. Solutions range from mandatory cycle annotations to implicit rules.

A designer using Ocapi specifies state machines and each state gets a cycle. State machines in the SpecC refinement flow may start with implicit clock boundaries, but they are made concrete eventually. SystemC's combinational processes become combinational logic, but its sequential processes denote cycle boundaries with *wait* calls.

Typical in high-level synthesis, HardwareC supports timing constraints such as "these three statements must execute in two cycles." While such constraints can be subtle for the designer and challenging for the compiler, they allow easier design-space exploration. Bach C is similar.

The C2Verilog compiler inserts cycles using complex rules and provides mechanisms for imposing timing constraints. Unlike HardwareC, these constraints are outside the language.

Transmogrifier C and Handel-C use implicit rules for inserting clocks. In Handel-C, only assignment and *delay* statements take a clock cycle. In Transmogrifier C, only loop iterations and function calls take a cycle. While simple to understand, such rules can require recoding to meet timing. Handel-C may require assignment statements to be fused and loops may need to be unrolled in Transmogrifier C.